\documentclass{iaus}
\usepackage{graphicx}

\title[JD 11.~~Model atmospheres and stellar limb-darkening] 
{Comparison of limb-darkening laws from plane-parallel and spherically-symmetric model stellar atmospheres}

\author[Hilding Neilson]   
{Hilding R. Neilson$^1$
}

\affiliation{$^1$Argelander-Institut f\"{u}r Astronomie, Bonn Universit\"{a}t, \\ 
Auf Dem H\"{u}gel 71, Bonn, D-53121, Germany
 \\ email: {\tt hneilson@astro.uni-bonn.de}}

\pubyear{2011}
\volume{282}  
\jname{From Interacting Binaries to Exoplanets:
Essential Modeling Tools}
\editors{A.C. Editor, B.D. Editor \& C.E. Editor, eds.}
\begin{document}

\maketitle

\begin{abstract}
Limb-darkening is a fundamental constraint for modeling eclipsing binary and planetary transit light curves.  As observations, for example from  {\it{Kepler}}, {\it{CoRot}}, and {\it{Most}},  become more precise then a greater understanding of limb-darkening is necessary. However, limb-darkening is typically modeled as simple parameterizations fit to plane-parallel model stellar atmospheres that ignores stellar atmospheric extension.  In this work, I compute linear, quadratic and four-parameter limb-darkening laws from grids of plane-parallel and spherically-symmetric model stellar atmospheres in a temperature and gravity range representing stars evolving on the Red Giant branch. The limb-darkening relations for each geometry are compared and are found to fit plane-parallel models much better than the spherically-symmetric models. Assuming that limb-darkening from spherically-symmetry model atmospheres are more physically representative of actual stellar limb-darkening than plane-parallel models then these limb-darkening laws will not fit the limb of a stellar disk leading to errors in a light curve fit.  This error will increase with a star's atmospheric extension.
\end{abstract}

\firstsection 
\section{Introduction}
Stellar limb-darkening is the observed change of intensity from the center of the stellar disk to the observable edge, where the intensity decrease is due to the geometric projection of the line-of-sight relative to the radius of the star. This effect is an important challenge for the interpretation of observations of binary stars (e.g.~\cite[Claret 2008]{Claret2008}), and planetary transits (e.g.~\cite[Knutson et al.~2007, Croll et al.~2011]{Knutson2006,Croll2011}), as well as interferometric (e.g.~\cite[Wittkowski et al.~2004]{Wittkowski2004}) and microlensing (e.g.~\cite[Zub et al.~2011]{Zub2011}) observations. Typically, limb-darkening is treated as a parameterization or relation as a function of the cosine of the angle formed by the radius and line-of-sight, called $\mu$ to simplify analysis. 

Stellar atmosphere models and binary/transit observations are complementary tools for understanding limb-darkening and stellar astrophysics in general because observed limb-darkening can help constrain models.   There are a plethora of articles describing different limb-darkening relations (\cite[Al-Naimiy 1978, Wade \& Ruchinski 1985, Claret et al.~1995, Claret~2000]{Al-Naimiy1978,Wade1985,Claret1995,Claret2000}), limb-darkening coefficients from predicted intensity profiles for a number of stellar atmosphere codes such as \textsc{Atlas}, and \textsc{Phoenix} (e.g. \cite[Howarth 2011, Sing 2010, Claret \& Hauschildt 2003]{Howarth2011, Sing2010, Claret2003}, and different fitting methods (\cite[Wade 1985, Heyrovsky 2003, 2007, Claret 2008]{Heyrovsky2003, Claret2008, Heyrovsky2007, Wade1985}).  In this work, I focus on a small number of limb-darkening laws and compare predicted fits for intensity profiles from plane-parallel and spherically symmetric model stellar atmospheres.  In the next section, I describe the stellar atmosphere code and three limb-darkening laws of interest: a linear, quadratic, and four-parameter (\cite[Claret 2000]{Claret2000}).  In Sect.~3, I present results of the fitting the limb-darkening laws using model atmospheres, and how the errors of the fit depend on assumed geometry. I summarize this work in Sect.~4.

\section{Stellar atmosphere code and Limb-darkening Laws}
For this analysis, I use a new Fortran 90 version of the Kurucz \textsc{Atlas} code (\cite[Lester \& Neilson 2008]{Lester2008}).  The code computes opacities using opacity distribution functions, atmospheres are assumed to be in local thermodynamic equilibrium, and hydrostatic equilibrium.  Each atmosphere model outputs intensity profiles as a function of wavelength, for an equal spacing of $\mu$ for  $1000$ points.  Typical calculations for the \textsc{Atlas} code compute intensity profiles for 10 - 17 $\mu$-points. The program computes models for either plane-parallel or spherically-symmetric geometries, where the plane-parallel model is described by two fundamental parameters such as $T_{\rm{eff}}$ and $\log g$, while spherical models require an additional parameter such as stellar mass.  \cite[Neilson \& Lester (2008)]{Neilson2008} fit model intensity profiles to interferometric observations from \cite[Wittkowski et al.~(2004)]{Wittkowski2004} and predicted similar fundamental parameters as those authors.  Also, \cite[Neilson \& Lester (2011)]{Neilson2011} predicted limb-darkening coefficients for a specific limb-darkening law from spherical models and compared to results for microlensing observations from \cite[Fields et al.~(2003)]{Fields2003} and found better agreement than the authors did using plane-parallel models and spherical models that had intensity profiles clipped to remove the extended limb.

I have computed approximately 2000 model stellar atmospheres in spherical symmetry for the parameter range $T_{\rm{eff}} = 3000~$-$~8000~$K in steps of $100~K$, $\log g=  -1~$-$~+3$ in steps of $0.25$ in cgs units, and $M = 2.5~$-$~10~M_\odot$ in steps of $2.5~M_\odot$.  Plane-parallel models are computed for the same values of $T_{\rm{eff}}$ and $\log g$.  

For this work, I compute least-squared fits to three laws:
\begin{eqnarray} 
\frac{I(\mu)}{I(1)} & = &1 - a(1-\mu) \mbox{\hspace{2.8cm}{Linear,}} \\
\frac{I(\mu)}{I(1)} & = &1 - b(1-\mu) - c(1-\mu)^2 \mbox{\hspace{1cm}{Quadratic,}} \\
\frac{I(\mu)}{I(1)} & = &1 - \sum_{i = 1}^{4} d_i(1-\mu^{i/2}) \mbox{\hspace{1.75cm}{Four Parameter,} }
\end{eqnarray}
where intensities are computed in the {\it{Kepler}} white light passband.  All fits are computed using least-square fitting of the limb-darkening coefficients.  The quality of the fit may be measured in a number of ways; here, I test the quality of the fit of limb-darkening laws by checking how well they conserve stellar flux, $\Delta F/F = (F_{\rm{Model}} - F_{\rm{Law}})/F_{\rm{Model}}$.

\section{Results \& Summary}
\begin{figure}[t]
\begin{center}
 \includegraphics[width=2.4in]{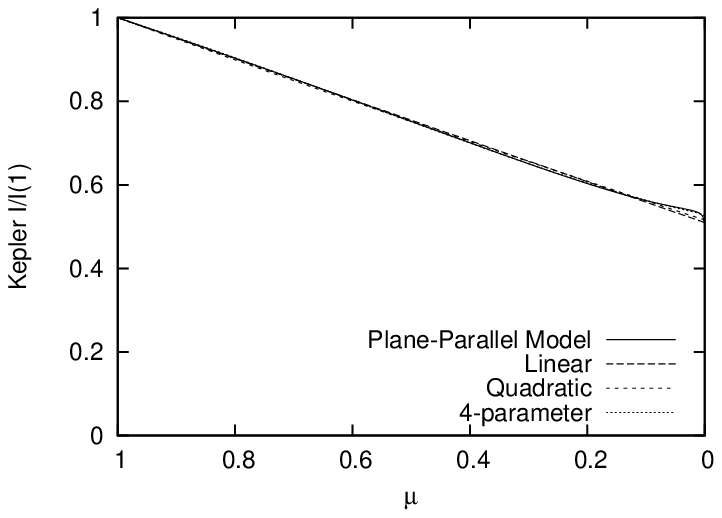} \includegraphics[width=2.4in]{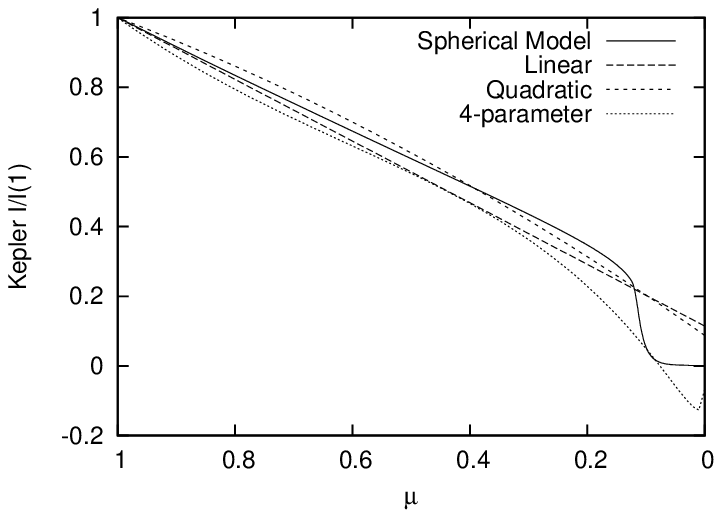} 
 \caption{Intensity profiles for a plane-parallel (left) and spherically-symmetric (right) model atmosphere with $T_{\rm{eff}} = 4000$~K and $\log g = 2$, along with the best-fit limb-darkening laws. }
   \label{f0}
\end{center}
\end{figure}

In Fig.~\ref{f0}, I show the predicted intensity profiles for a $T_{\rm{eff}} = 4000~$K and $\log g = 2$ model atmosphere for both geometries.  There is a significant difference between the model intensity profiles such that the intensity near the limb of the spherically-symmetric model atmosphere is much smaller than the plane-parallel model atmosphere.  The plane-parallel model does not appear to go to zero, though the equation of transfer suggests that the intensity in the limit as $\mu \rightarrow 0$ then $I(\mu) \rightarrow 0$ (\cite[Mihalas 1978]{Mihalas1978}). This may be an issue with the resolution of $\mu$.  The plane-parallel model is also much better fit by the limb-darkening laws than the spherically-symmetric model atmosphere because the spherically-symmetric intensity profile is more complex.

\begin{figure}[t]
\begin{center}
 \includegraphics[width=2.4in]{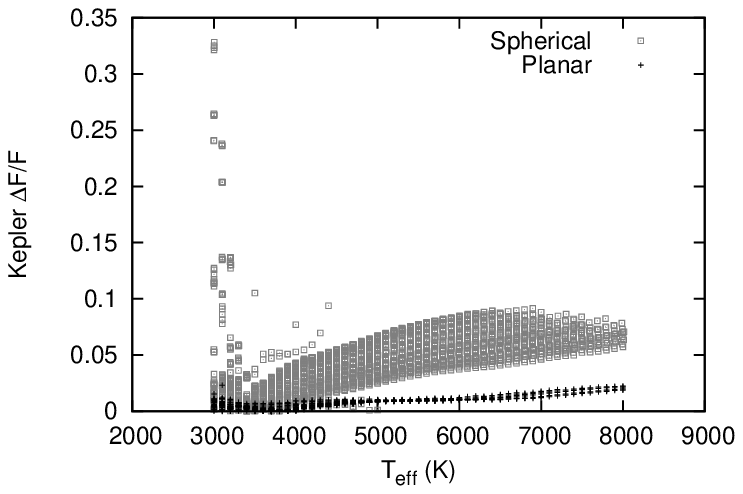} \includegraphics[width=2.4in]{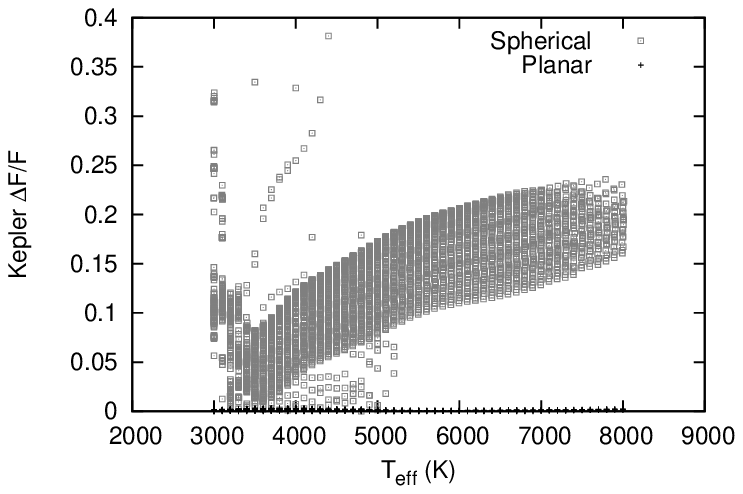}
 \includegraphics[width=2.4in]{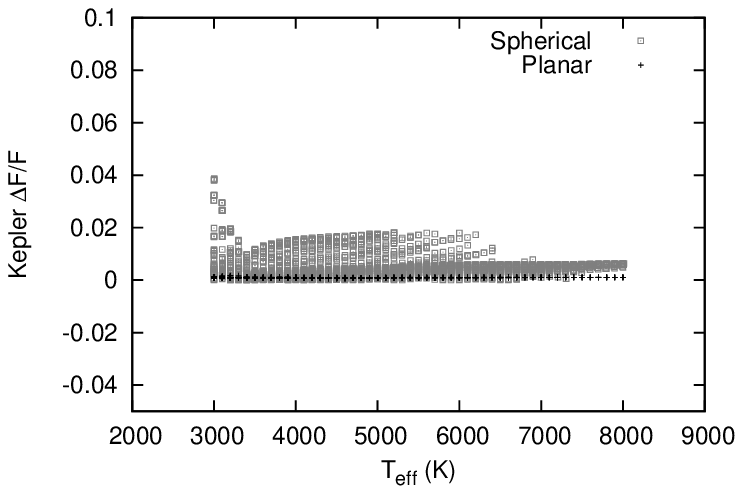} 
 \caption{Relative difference of between model and fit stellar fluxes for the three limb-darkening law in the {\it{Kepler}} passband, linear (upper-left), quadratic (upper-right), and \cite[Claret (2000)]{Claret2000} four-parameter (bottom) laws for plane-parallel and spherically-symmetric model stellar atmospheres.}
   \label{fig1}
\end{center}
\end{figure}

The relative difference between the model intensity profiles and limb-darkening laws predicted stellar fluxes are shown in Fig.~\ref{fig1}.  The fits to plane-parallel model atmospheres appear to be better, the average error is $< 5\%$ for the linear law, and is much smaller for the other laws.  The quality of the fit for plane-parallel model atmospheres  is also apparently independent of effective temperature.  The results for the spherically-symmetric model atmospheres are strikingly different.  The flux errors are much larger, $5$-$10\%$ for the linear law, $0$-$20\%$ for the quadratic law, and $0$-$5\%$ for the four parameter limb-darkening law.  The difference in fits due to model atmosphere geometry suggests a significant problem for understanding stellar limb-darkening.  It is reasonable to assume that a spherical geometry is a more physical representation of an actual star than a plane-parallel stellar atmosphere, hence people should be hesitant for using limb-darkening coefficients generated from plane-parallel model atmospheres. It also suggests that these are not ideal limb-darkening laws to use and it may be necessary to develop new limb-darkening relations for the future. 

\nocite{Claret2008}
\nocite{Knutson2006}
\nocite{Croll2011}
\nocite{Wittkowski2004}
\nocite{Zub2011}
\nocite{Mihalas1978}
\nocite{Al-Naimiy1978}
\nocite{Wade1985}
\nocite{Claret2000}
\nocite{Claret1995}
\nocite{Howarth2011}
\nocite{Sing2010}
\nocite{Claret2003}
\nocite{Heyrovsky2007}
\nocite{Heyrovsky2003}
\nocite{Lester2008}
\nocite{Neilson2008}
\nocite{Gustafsson2008}
\nocite{Hauschildt1999}
\nocite{Neilson2011}
\nocite{Fields2003}

\begin{discussion}
\discuss{R.E. Wilson}{Your starting explanation of why the intensity goes to zero at the limbs for a plane parallel case is not correct.  Actually the intensity does not go to zero at the limb.  The mistake came from neglect of emission along the line of sight Ð it is not just an attenuation problem, but has both a source factor and an attenuation factor in the intensity integral.  If one looks into an infinite uniform region, the received intensity is not zero, but is the intensity characteristic of the regionÕs temperature.}

\discuss{I. Hubany}{To Bob Wilson's comment: The intensity does not indeed have
to go to zero at the limb, but such a case is not covered by a
1-D plane-parallel treatment of the transfer equation anyway because
in this case the medium is infinite with no natural boundary condition.
Comment to the talk: The Eddington factor (the ratio of the K-moment
and the mean intensity) is not necessarily equal to 1/3 in the
plane-parallel case. Such quantity is usually called a variable
Eddington factor, and depends on depth and frequency; it goes to
1/3 only deep in the atmosphere.}

\discuss{A. Pr\v{s}a}{I understand why one would use an analytic model for a Mandel-Agol type approach, but perhaps the systematic error from the simple fit may be avoided simply by linearly interpolating (or looking up) $I(\mu)$.}

\discuss{Ph. Stee}{I did not understand why you used the first lobe of the visibility function to fit the LD instead of the second lobe, especially since you may fit the first lobe with a uniform disk?}
\end{discussion}
\end{document}